\documentclass[12pt]{article}
\usepackage{hhline}
\usepackage{latexsym,graphicx}
\usepackage{subfigure}
\usepackage{amssymb}
\usepackage{amsmath}
\usepackage{amscd}
\usepackage{amsthm}
\usepackage{float}
\usepackage{xcolor}
\usepackage[left=2cm,top=2.5cm,right=2.5cm,bottom=1.5cm]{geometry}   
 
\begin{document}
\begin{center}
\large{\bf{ Big rip in Swiss-cheese Brane-worlds with cosmic transit}} \\
\vspace{10mm}
\normalsize{Nasr Ahmed$^1$'$^3$, M. Fekry$^2$ \& Tarek M. Kamel$^3$ }\\
\vspace{5mm}
\small{\footnotesize $^1$ Mathematics and statistics Department, Faculty of Science, Taibah University, Saudi Arabia.} \\
\small{\footnotesize $^2$ Department of Basic Science, Common First Year, King Saud University, Saudi Arabia} \\
\small{\footnotesize $^3$ Astronomy Department, National Research Institute of Astronomy and Geophysics, Helwan, Cairo, Egypt\footnote{abualansar@gmail.com, Tel: +966583689312}} 
\end{center}  
\date{}
\begin{abstract}
We study the big rip scenario in Swiss-cheese Brane-worlds. The results obtained have been found to be independent of the value of the cosmological constant $\Lambda$ whether it is positive, negative or zero. Negative tension branes are not allowed in the current model. There is a sign flipping in cosmic pressure corresponding to the sign flipping in the deceleration parameter from positive to negative. The evolution of the EoS parameter shows the presence of three phases: the matter dominant decelerating era, the accelerated-Quitessence phase, and the phantom phase. The evolution of the potential and Kinetic term also shows a change of sign. The energy conditions and cosmographic parameters have also been investigated.
\end{abstract}

\section{Introduction and motivation}

The mysterious dark energy with negative pressure has been introduced as a possible explanation to the accelerating cosmic expansion \cite{11,13,14}. Some dynamical scalar fields models of dark energy have been proposed such as quintessence \cite{quint}, phantom energy \cite{phant},tachyon \cite{tak}, k-essence \cite{ess}, Chaplygin gas \cite{chap}, holographic \cite{holog1, holog2} and ghost condensate \cite{ark,nass}.  Modifying the geometrical part of the Einstein-Hilbert action is another approach \cite{moddd}. Examples of modified gravity theories are $f(R,T)$ gravity where $R$ is the Ricci scalar and $T$ is the trace of the energy momentum tensor\cite{nnn1}, Gauss-Bonnet gravity \cite{noj8}, $f(T)$ gravity \cite{torsion} where $T$ is the torsion scalar, Mimetic gravity \cite{sunny}, a good review has been given in \cite{add1}. Brane-worlds \cite{brane} represent an interesting extra-dimensional modified gravity where the universe is a $3 + 1$ -dimensional surface (brane) embedded in a $3 + 1 + d$ -dimensional space-time (bulk). The idea of a domain-wall universe was first introduced in 1983 \cite{brane1} while the effective gravitational equations on the brane have been derived in \cite{brane3}. The modified Einstein equations on the brane can be written as:
\begin{equation}
G_{ab}=-\Lambda g_{ab}+\kappa^2T_{ab}+\tilde{\kappa}^4S_{ab}-\varepsilon_{ab},
\end{equation}
$G_{ab}$ is the Einstein tensor, $g_{ab}$ is the metric tensor and $\varepsilon_{ab}$ is the electric part of the Weyl curvature of the bulk. $S_{ab}$ is a quadratic expression in the energy momentum tensor $T_{ab}$:
\begin{equation}
S_{ab}=\frac{1}{12}TT_{ab}-\frac{1}{4}T_{ac}T^c_b+\frac{1}{24}g_{ab}(3T_{cd}T^{cd}-T^2),
\end{equation}
$\kappa^2$ and $\Lambda$ are the brane gravitational and cosmological constants respectively. The (positive) brane tension $\lambda$, the bulk cosmological constant $\tilde{\Lambda}$, and the bulk gravitational constant $\tilde{\kappa}^2$ are related by 
\begin{eqnarray}
6\kappa^2 &=&\tilde{\kappa}^2\lambda,\\
2\Lambda &=& \kappa^2\lambda+\tilde{\kappa}^2 \tilde{\Lambda}.
\end{eqnarray}

While negative tension branes are unstable, the universe can also be described  using a variable brane tension \cite{brane7}. Examples of brane-world models are the DGP model \cite{brane6ii}, the thick brane model \cite{brane6iii}, GRS model \cite{brane6i}, and the universal extra dimension model \cite{yours}. Brane-world cosmology has some interesting features such as the phantom behavior with no big rip singularity and self-acceleration of geometrical origin \cite{brane6iiii4}. Brane-worlds in modified $f(R, T)$ gravity have been studied in \cite{moreas}. The effect of one extra dimension in brane-world cosmology has been investigated in \cite{brane10}.\par

Another interesting brane-world cosmology has been discussed in \cite{brane14} in which black strings in the higher-dimensional bulk penetrate the lower-dimensional brane \cite{add4, brane15,mypaper}. Black strings are considered as the simplest higher-dimensional extensions of black holes. This penetration gives rise to Schwarzschild black holes immersed in the FLRW brane which leads to a brane with 'Swiss-cheese' structure. Such a model could give a more realistic cosmological description than the FLRW brane because of the existence of Schwarzschild black holes and can lead to important cosmological consequences. Since black holes are not isolated objects, they should be included in any realistic cosmological model \cite{br50}. This cosmic Swiss-cheese idea was introduced for the first time by Einstein and Strauss in 1945 as a method of embedding a compact object within the universe \cite{br51,br52}. The method is based on removing a spherical comoving region from a FLRW space-time and replacing it by a point mass at the its center which creates a hole. Having several non-overlapping spheres where the mass of each sphere contracted to the center (Schwarzschild black holes and surrounding vacuum regions) creates inhomogeneities. The Einstein-Strauss Swiss-cheese inhomogeneous cosmological model has predicted a modified Luminosity-redshift relation \cite{br53}. \par

The main purpose behind Swiss-cheese cosmology is to construct locally inhomogeneous space-times which appear globally isotropic and satisfy Einstein equations. Since cosmological observations depend mainly on photons as the carriers of information, understanding how light propagates through the universe is essential for a correct interpretation of observational data. In the standard FLRW model, light propagates through a FLRW space-time which describes a perfectly homogenous and isotropic universe. Since the real universe is not homogeneous, some alternative models for the inhomogeneous universe have been suggested such as the Swiss-cheese models and the lattice models \cite{br54}. Instead of constructing a specific space-time model, and inspired by Zel’dovich original idea \cite{br55}, a different approach has been introduced by dyer and Roeder \cite{br56,br57}. The original approximation suggested by Zel’dovich considers the propagation of light through emptier rather than denser cosmic regions and is now known as the Dyer–Roeder approximation. Both the Dyer-Roeder approximation and Swiss-cheese models have been shown to be indistinguishable regarding the interpretation of cosmological data, they both generate the same distance-redshift relation \cite{br58}.\par
In Swiss-cheese brane-world scenario, our $4$D  universe is regarded as a FLRW brane moving in a static $5$D Schwarzschild anti-de Sitter bulk. The FLRW metric given by
\begin{equation}
ds^{2}=-dt^{2}+a^{2}(t)\left[ \frac{dr^{2}}{1-Kr^2}+r^2d\theta^2+r^2\sin^2\theta d\phi^2 \right] \label{RW}
\end{equation} 
Considering only the flat case ($K=0$) supported by observations \cite{flat1, flat2, flat3}, the cosmological equations are
\begin{eqnarray}
\frac{\dot{a}^2}{a^2} &=& \frac{\Lambda}{3}+\frac{\kappa^2 \rho}{3}\left(1+\frac{\rho}{2\lambda}\right),  \label{cosm1}\\
\frac{\ddot{a}}{a} &=& \frac{\Lambda}{3}-\frac{\kappa^2 }{6}\left[\rho\left(1+\frac{2\rho}{\lambda}\right)+3p \left(1+\frac{\rho}{\lambda}\right) \right] .\label{cosm2}
\end{eqnarray}
General Relativity can be recovered for $\rho/\lambda\rightarrow 0$. All results in the current work have been found to be $\Lambda$-independent, the analysis is the same for any value of the cosmological constant whether it is positive, negative or zero. We have also tested some models for the time-dependent $\Lambda$ such as \cite{20g, 20n} $\Lambda(t) =  \frac{\Lambda_{Pl}}{\left(t/t_{Pl}\right)^2} \propto \frac{1}{t^2}$, $\Lambda(H)= \beta H +3H^2 + \delta H^n \,\, (n \in R-\left\{0,1\right\}$) and $\Lambda(H, \dot{H}, \ddot{H})=\alpha+\beta H+\delta H^2+ \mu \dot{H}+\nu \ddot{H}$ with $H$ is the Hubble parameter. They all leaded to the same results. 
The paper is organized as follows: In section 2, we study the solution and its physical properties. The cosmographic analysis has been provided in section 3. The final conclusion is included in section 4.

\section{The Big rip solution}

The phantom energy is characterized by an energy density that increases with time and a super-accelerated expansion. If a scalar field $\phi$ describes dark energy with the FRW customary definitions $\rho=\dot{\phi}^2/2+V(\phi)$ and $p=\dot{\phi}^2/2-V(\phi)$ where $V(\phi)$, $\rho$ and $p$ are the field potential, energy density and pressure respectively. Then, the kinetic term $\dot{\phi}^2/2 < 0$ and, consequently, the phantom cosmology faces severe instabilities and classical inconsistencies \cite{2br}. The dominant energy condition will also be violated such that $ \rho +p < 0$, and the future big rip singularity will appear near of which the universe may suffer from causality violations.\par 

The idea of a phantom dominated super-accelerated expanding universe with $\omega<0$ leads to a finite-time future singularity called the Big Rip. In Such scenario with a divergent scale factor, and due to the super-exponential expansion, all the structures of the universe might be ripped apart \cite{rip1,rip2}. As the cosmic time $t$ approaches the Big Rip singularity $t \rightarrow t_s$, the scale factor $a \rightarrow \infty$, the energy density $\rho \rightarrow \infty$, the pressure of the dark energy $|p| \rightarrow \infty$, the Hubble parameter $H \rightarrow \infty$ and its derivative $\dot{H} \rightarrow \infty$. The dark energy component with super-negative equation of state has been shown to be allowed by some observational data \cite{phant}. It is also widely believed that this classically predicted singularity may be avoided in quantum gravity. The question of how the big rip can be avoided in the far future has been addressed by many authors \cite{rip3,rip4,rip5,rip6}. The classical and quantum occurrence of the big rip in the framework of $f(R)$ modified gravity has been discussed in \cite{rip5}. In the context of general relativity, it has been shown that the big rip could be avoided due to the gravitational Schwinger pair-production \cite{rip7}. In a scalar model with Kinetic and Gauss-Bonnet couplings, it has been found that the big rip occurrence depends on the parameters of the solution \cite{rip8}.

Several scale factors for the big rip singularity have been used in the literature. The Hubble parameter $H=\frac{h_0}{t_0-t}$ which leads to the scale factor $a(t)=C(t_0-t)^{-h_0}$, where $h_0$ and $t_0$ are positive constants, has been considered in \cite{brfr} for the study of the big rip in $f(R)$ gravity. A scale factor for two big rips (a future one and a past one) has been introduced in \cite{2br} as $a(t)=\alpha(\beta + x \tan x)$ where $\kappa$, $\alpha$ and $\beta$ are positive constants and $x=\kappa t$. The ansatz $a(t)=\left(\frac{t}{t_s}\right)^q(a_s-1)+1-\left(1-\frac{t}{t_s}\right)^n$ has been used in \cite{borro} where the evolution of $a(t)$ occurs in the interval $0<t<t_s$ with $a_s \equiv a(t_s)$, $1<n<2$ and $0<q\leq 1$. It was shown in \cite{bri3} that this scale factor can be expanded at late times in powers of $(t_s-t)$ and the approximation $a\approx a_s +\frac{q(1-a_s)}{t_s}(t_s-t)$ has been used.
In the current work we consider the scale factor
\begin{equation}  \label{ss}
a(t)=a_0\left(\frac{t}{t_s-t}\right)^n
\end{equation}
where $a_0$, $t_s$ and $n$ are positive constants, the BR singularity happens at $t=t_s$. The scale factor (\ref{ss}) leads to a Hubble parameter and a linearly varying deceleration parameter DP in the form
\begin{equation}  \label{dp}
H(t)=-\frac{n t_s}{t(t-t_s)}~~~~~~,~~~~~~q(t)=-\frac{1}{nt_s} \left(2t+t_s(n-1)\right)
\end{equation}
a time-dependent deceleration parameter (DP), in general, provides a more realistic description to cosmic evolution than a constant one. A constant DP, such as Berman's model suggested in \cite{berman}, doesn't agree with the modern picture of cosmic evolution and cosmic transit where the DP changes sign with different eras \cite{dppaper}. The DP evolves from $1/2$ to $-1$ in the standard $\Lambda CDM$ model. A more general approach to expand the DP in Taylor series has been considered in \cite{dppaper} as
\begin{equation}  
q(x)= q_0+q_1\left(1-\frac{x}{x_0}\right)+q_2\left(1-\frac{x}{x_0}\right)^2+...
\end{equation}
where $x$ can be the scale factor $a$, the redshift $z$, cosmic time etc. The first two terms only represent a linear approximation. Using the scale factor (\ref{ss}) with the cosmological equations (\ref{cosm1}) and (\ref{cosm2}) leads to two different solutions for the energy density $\rho$ and one solution for the pressure $p$ . 
	\begin{equation}
\begin{array}{l}
\rho (t) = \frac{1}{2}\frac{{ - 2\pi t^2 \lambda  + 2\pi t\lambda t_s  + \sqrt {4\pi ^2 t^4 \lambda ^2  - 8\pi ^2 t^3 \lambda ^2 t_s  + 4\pi ^2 t^2 \lambda ^2 t_s^2  - \pi t^4 \Lambda \lambda  + 2\pi t^3 \Lambda \lambda t_s  - \pi t^2 \Lambda \lambda t_s^2  + 3\pi n^2 \lambda t_s^2 } }}{{\pi \left( { t - t_s } \right)t}} \\ 
	~~~~~~,  - \frac{1}{2}\frac{{2\pi t^2 \lambda  - 2\pi t\lambda t_s  + \sqrt {4\pi ^2 t^4 \lambda ^2  - 8\pi ^2 t^3 \lambda ^2 t_s  + 4\pi ^2 t^2 \lambda ^2 t_s^2  - \pi t^4 \Lambda \lambda  + 2\pi t^3 \Lambda \lambda t_s  - \pi t^2 \Lambda \lambda t_s^2  + 3\pi n^2 \lambda t_s^2 } }}{{\pi \left( { t - t_s } \right)t}} ~.  \\ 
	 p(t) = \frac{{4\rho (t)\pi t^2 \lambda t_s^2  - 8\rho (t)\pi t^3 \lambda t_s  + 4\rho (t)\pi t^4 \lambda  + t^2 \Lambda \lambda t_s^2  - 2t^3 \Lambda \lambda t_s  + t^4 \Lambda \lambda  - 3n^2 \lambda t_s^2  + n\lambda t_s^2  - 2n\lambda t_s t}}{{4t_s^2 \lambda \pi t^2  + 4t_s^2 \rho (t)\pi t^2  - 8t_s \lambda \pi t^3  - 8t_s \rho (t)\pi t^3  + 4\lambda \pi t^4  + 4\rho (t)\pi t^4 }} ~.  \\ 
\end{array}
	\end{equation}
We choose only the physically acceptable solution for $\rho$ where it is positive and it tends to $\infty$ as $t \rightarrow 0$ (investigating how negative energy density would affect a classic Friedmann cosmology has been done in \cite{negativerho}). The evolution of $\rho$, $p$ and the EoS parameter $\omega=\frac{p}{\rho}$ is shown in figure 1. Making use of the relation $a=\frac{1}{1+z}$, the time $t$ can be written in terms of $z$ as
\begin{equation}
t(z)=\frac{\left((1+z) a_0\right)^{-\frac{1}{n}} t_s}{\left((1+z) a_0\right)^{-\frac{1}{n}}+1}.
\end{equation}
Inserting this in the equation for $\omega(t)$, we get the EoS parameter expressed in terms of $z$. The evolution of $\omega(z)$ is shown in Figure 1, and we need to fine-tune between different parameters so that $\omega(z)=-1$ at $z \simeq 0$. Depending only on the value of $n$, $\omega(z)$ gets more closer to $-1$ at $z \simeq 0$. This is completely independent of the value of the cosmological constant $\Lambda$ and any positive values of the brane tension $\lambda$.\par

We see from the expression for $\rho$ that the energy density vanishes if $\lambda=0$, also negative values of $\lambda$ are not allowed where $\rho(t)$ will never be defined. This is a remarkable result where it has been shown in \cite{negativebr} that negative tension branes are unstable objects and even in the five-dimensional bulk there is no need to consider a negative tension brane. It is also clear that for $\rho(t)$ the denomerator $\pi(t-t_s)t =0$ at $t=t_s$ which means that the energy density diverges at the big rip. The evolution of cosmic pressure shows a sign flipping from positive to negative corresponding to transition from early-time decelerating epoch to late-time accelerating epoch.
 
While the pressure in the original Swiss-cheese brane-world model \cite{brane14} is positive with no dark radiation or cosmic transit where the cosmic expansion always decelerates for any value of $\Lambda$, the pressure in the current model shows a sign flipping from positive in  early decelerating time to negative in late accelerating time regardless of the value of $\Lambda$. The plots of $p$ and $q$ shows that the sign flipping of $p$ happens around same time where the deceleration parameter $q$ changes sign from positive to negative. This picture agrees with the dark energy assumption as a negative pressure component that causes a repulsive gravity effect. \par  

The evolution of the EoS parameter against cosmic time shows the presence of three phases: the matter dominant decelerating era above $\omega_{eff}=-1/3$, The accelerated-Quitessence phase between the line $\omega=-1/3$ and the phantom divide line $\omega=-1$, and the phantom phase below the line $\omega=-1$. Our results here agree with the work by Granda and Loaiza in \cite{rip8} where they Considered a scalar field model with kinetic and Gauss Bonnet couplings. Granda and Loaiza obtained solutions where it is possible to cross to the phantom era with or without big rip singularity. The evolution of the EoS parameter $\omega(t)$ in \cite{rip8} also goes through the same three phases. \par
The weak energy condition ”WEC” states that $\rho$ and $p$ obey the inequalities
\begin{eqnarray*}
\rho + p = \rho(1+\omega) \geq 0 \\ \nonumber
\rho \geq 0 
\end{eqnarray*}
Where $\rho + p <0$ if $\omega < -1$. For the current model we get ( for $n=0.25$ )
	\begin{eqnarray} \label{null}
	\rho(t)+p(t)=~~~~~~~~~~~~~~~~~~~~~~~~~~~~~~~~~~~~~~~~~~~~~~~~~~~~~~~~~~~~~~~~~~~~~~~~~~~~~~~~~~~~~~~~~~~~~~~~ \\   \nonumber
	\frac{\lambda n t_s(2t-t_s)(t-t_s)^{-1}}{2\sqrt{\pi} t\sqrt{\lambda(4\pi \lambda t^4-8\pi \lambda t^3t_s+4\pi \lambda t^2t_s^2-\Lambda t^4+2\Lambda t_s t^3- \Lambda t^2t_s^2+3n^2t_s^2)}}\left\{
        \begin{array}{ll}
            >0 , t \in (0,\frac{1}{2}] \\
            <0 , t \in [\frac{1}{2},1)
        \end{array}
    \right.
	\end{eqnarray}
And 
	\begin{equation}
	\rho(t)+3p(t)\left\{
        \begin{array}{ll}
            >0 , t \in (0,\frac{1}{2}] \\
            <0 , t \in [\frac{1}{2},1)
        \end{array}
    \right. ~0~~~,~~~~~~~~\forall t \in (0,1)~~~~~~~~~~~~~~~~~~~~~~~~~~~~~~~~~~~~~~~~~~~~~~~ \\   \nonumber
	\end{equation}
So, there is a violation for the null and strong energy conditions for $t \geq \frac{1}{2}$ where $\omega$ becomes less than $-1$. While these classical  linear energy conditions can not be valid in completely general situations \cite{parc}, and due to the existence of a quadratic energy term in the current model, we should also test the following new nonlinear energy conditions \cite{ec,FEC1,FEC2}:  (i) The flux energy condition $\rho^2 \geq p_i^2$ (ii) The determinant energy condition $ \rho . \Pi p_i \geq 0$ . (iii) The trace-of-square energy condition $\rho^2 + \sum p_i^2 \geq 0$ . For the current model we found that none of these three nonlinear energy conditions is satisfied. 
\subsection{The potential and the kinetic term}
\begin{figure}[H] \label{tap1}
  \centering             
  \subfigure[$a$]{\label{F1}\includegraphics[width=0.29\textwidth]{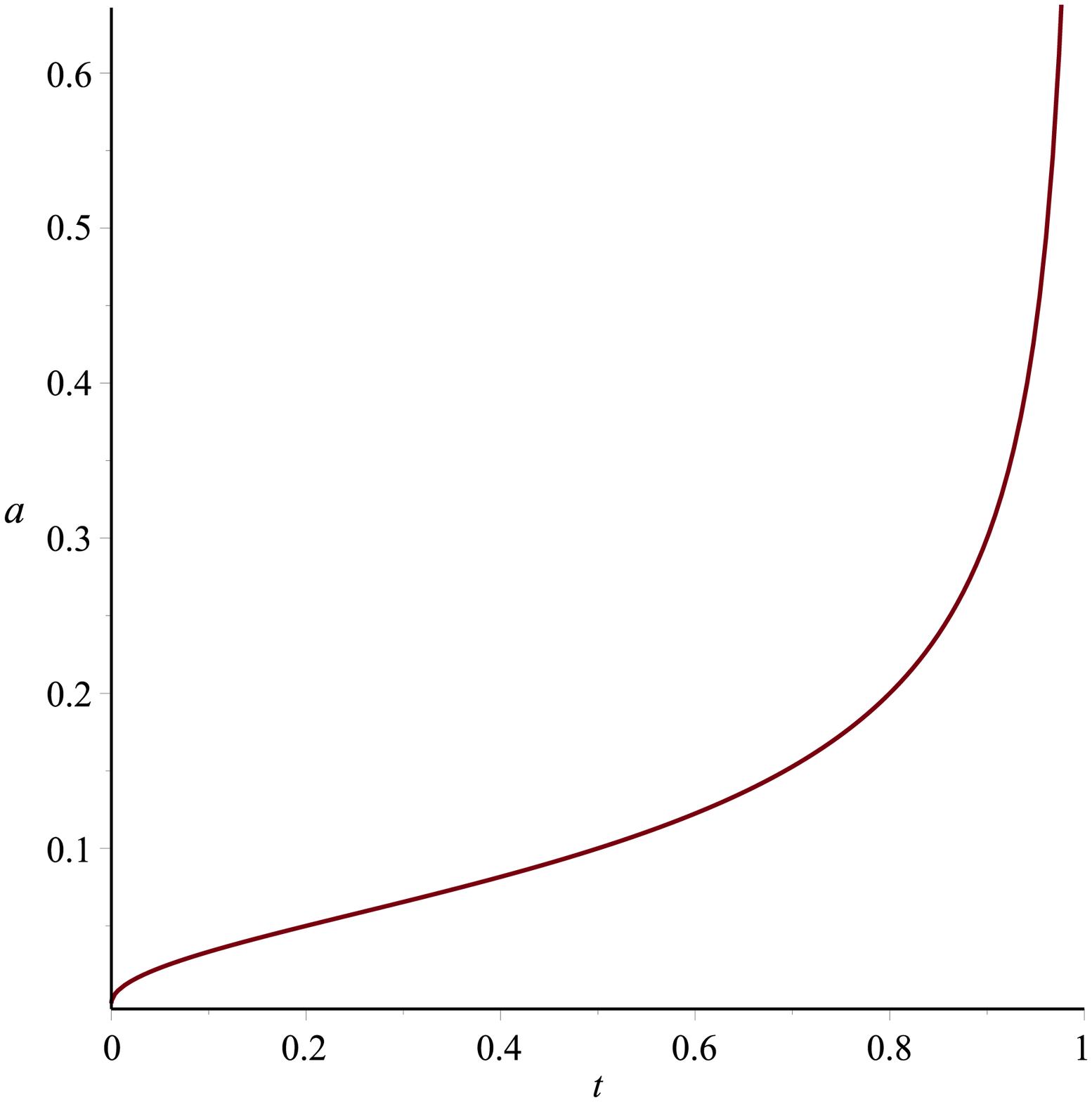}} 
			  \subfigure[$p$]{\label{F4}\includegraphics[width=0.29 \textwidth]{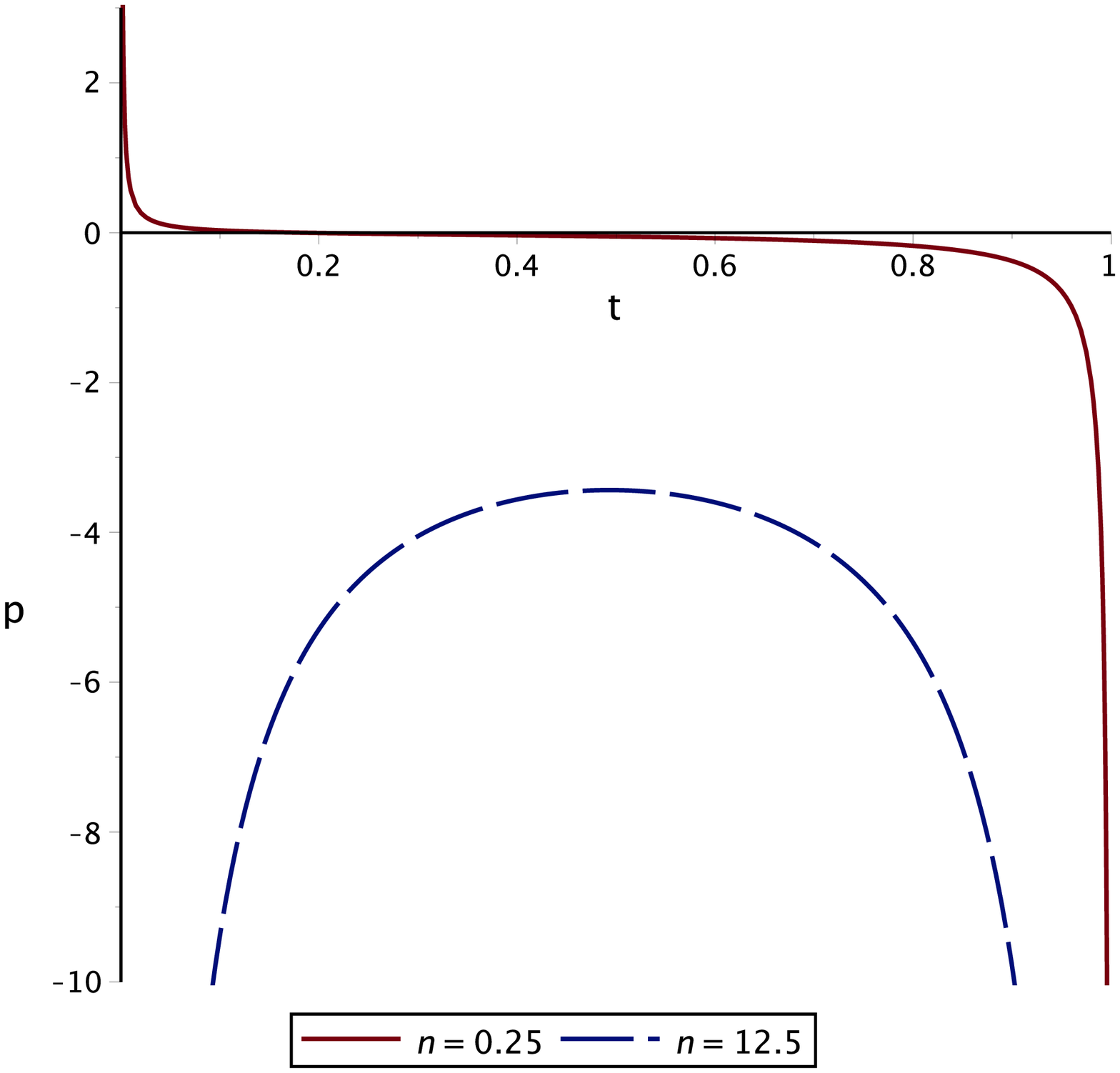}}
		\subfigure[$\rho$]{\label{F5}\includegraphics[width=0.29 \textwidth]{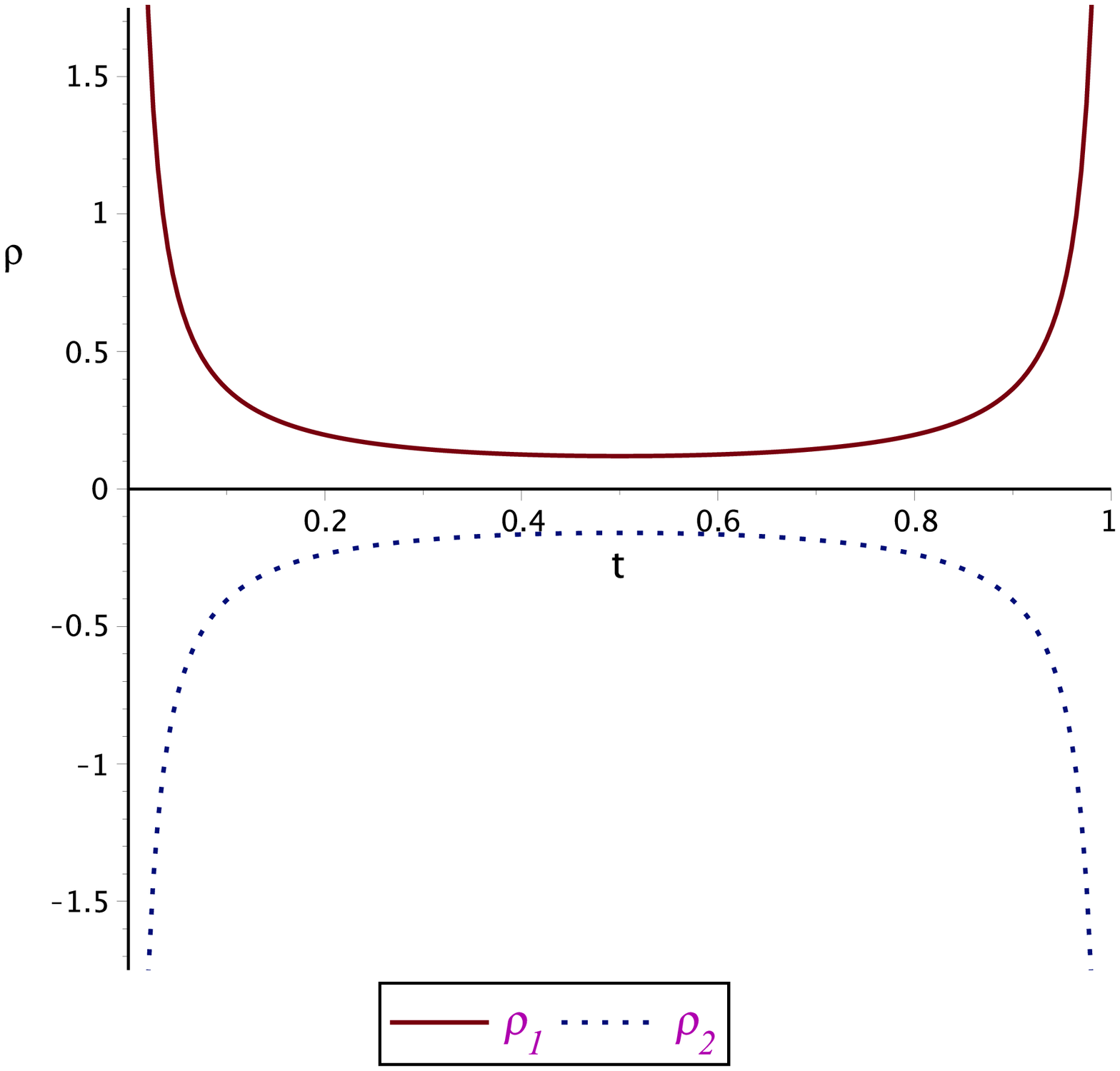}}\\
    \subfigure[$\omega(z)$]{\label{F9}\includegraphics[width=0.29\textwidth]{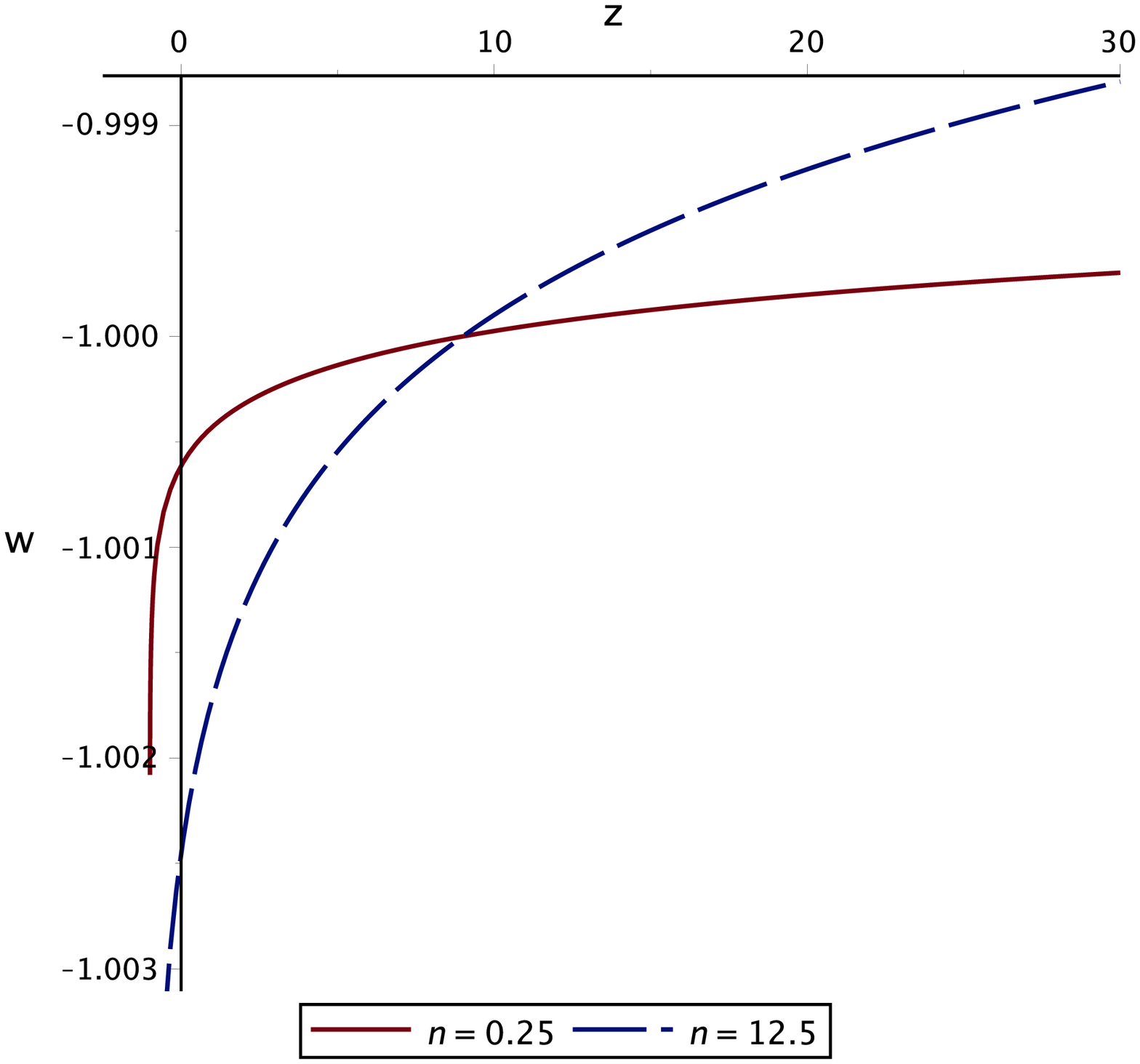}} 
				\subfigure[$\omega(t)$]{\label{F33}\includegraphics[width=0.29\textwidth]{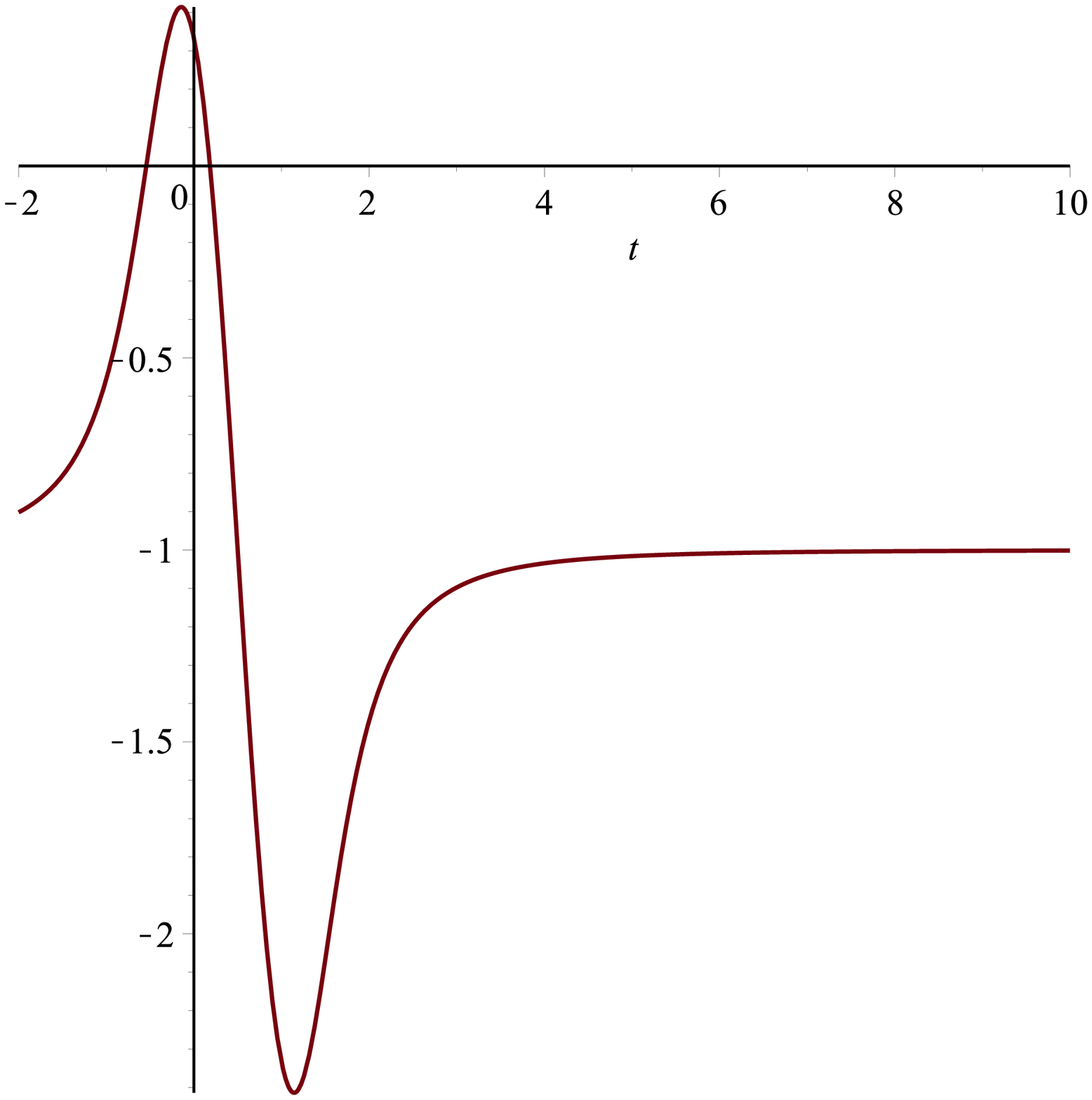}} 
					\subfigure[$V(t)$]{\label{F3522}\includegraphics[width=0.29\textwidth]{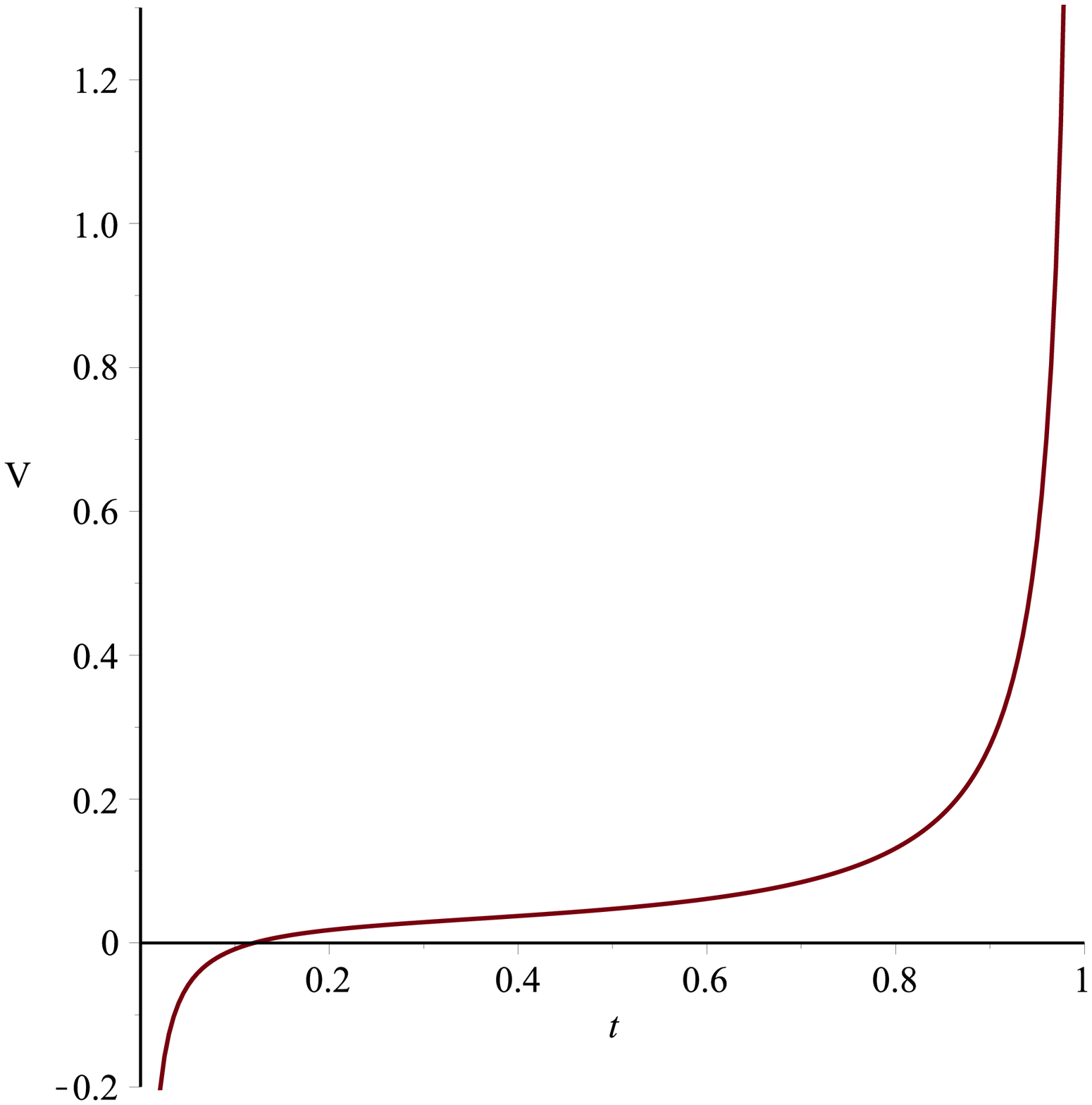}} \\
					\subfigure[$K(t)$]{\label{F350}\includegraphics[width=0.29\textwidth]{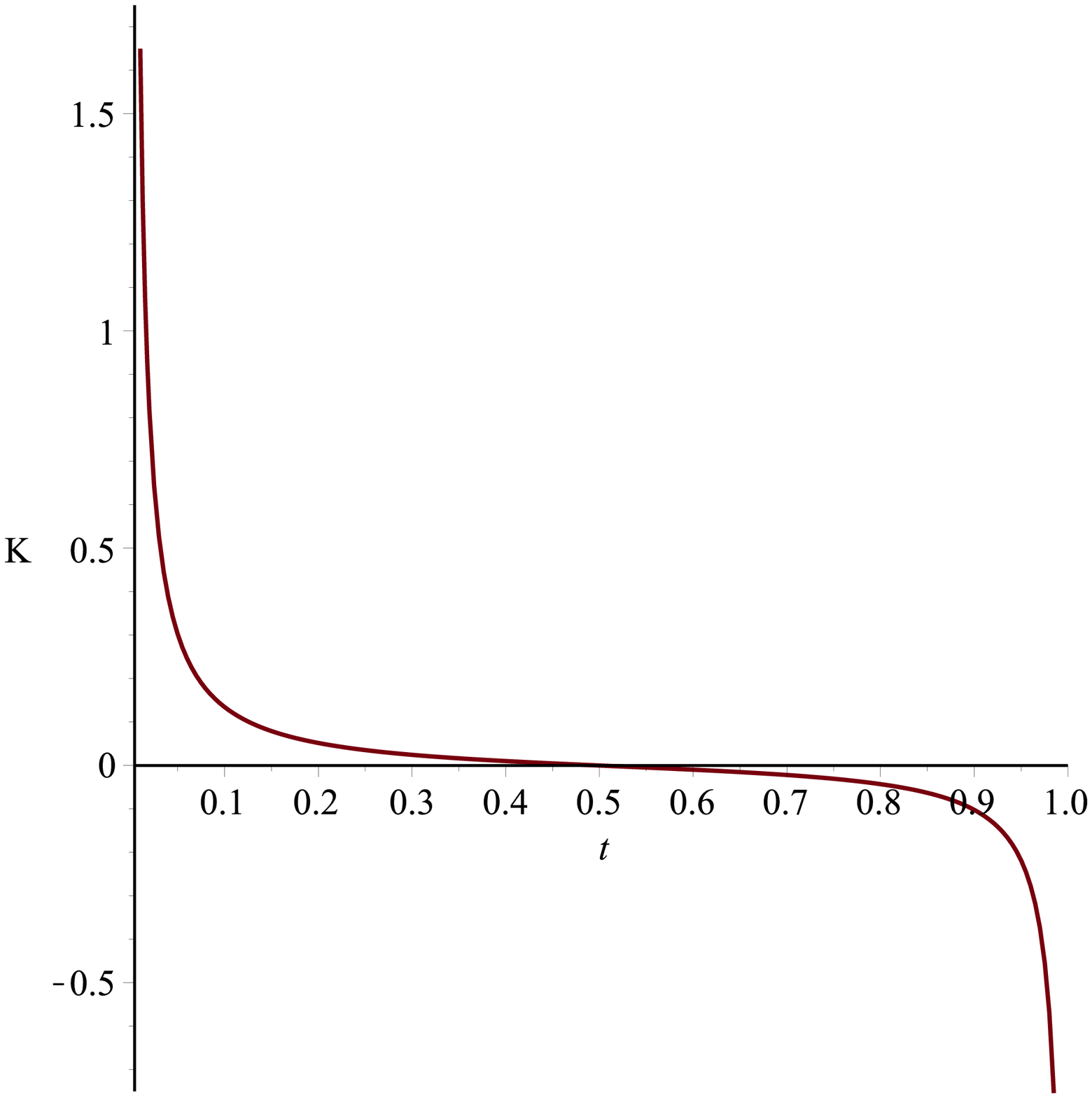}}
					\subfigure[$\phi(t)$]{\label{F359}\includegraphics[width=0.29\textwidth]{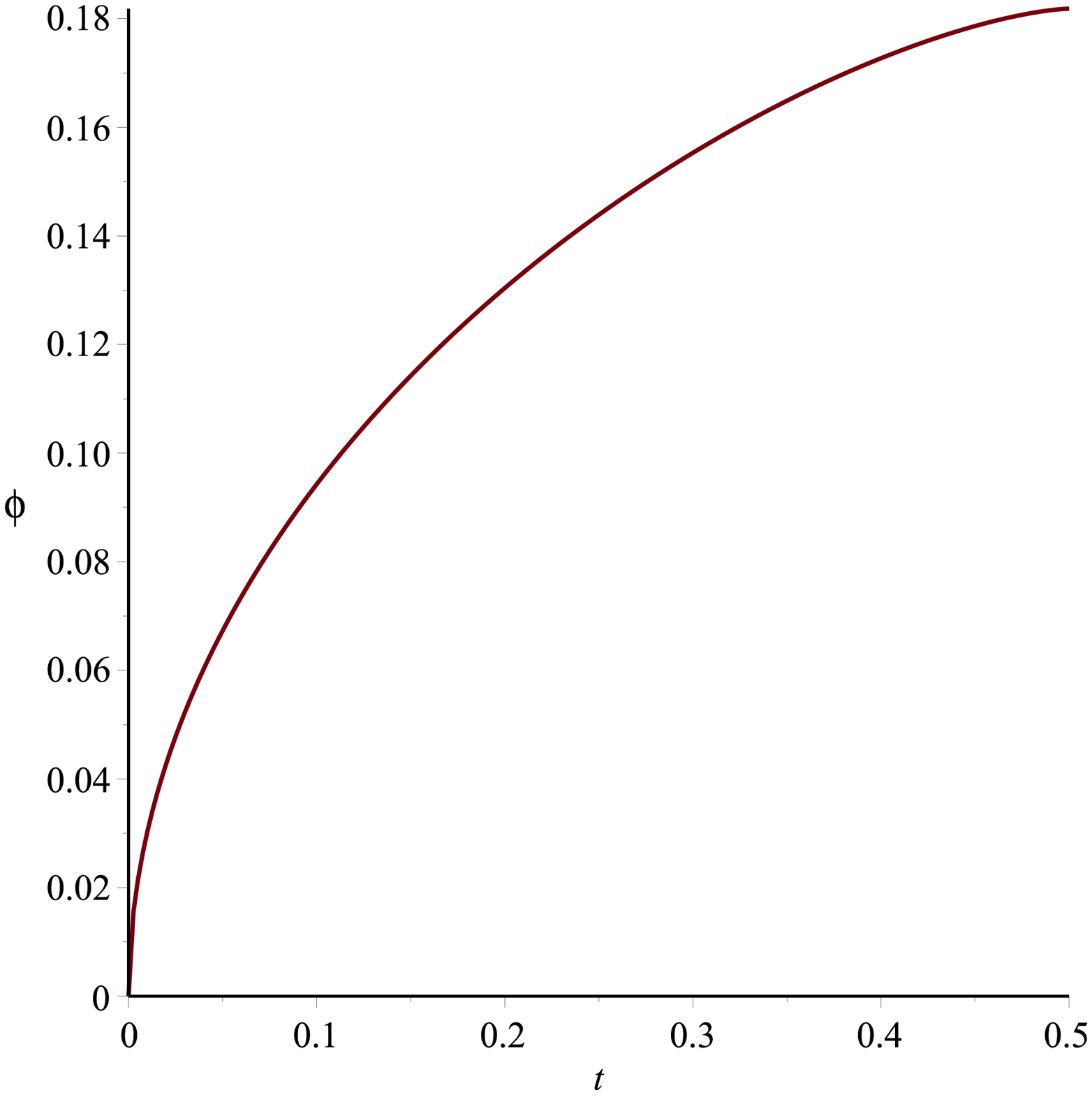}}
  \caption{ (a) The BR scale factor $a(t)$. (b) pressure goes from positive to negative. (c) energy density increases as the universe expands when $\omega <-1$. (d) $\omega(z) \simeq-1$ at $z= 0$. (e) $\omega(t)$ crosses the phantom devide line. (f) The scalar field potential $V=\frac{1}{2}(\rho-p)$. (g) The kinetic term $K=\frac{1}{2}\dot{\phi}^2=\frac{1}{2}(\rho+p)$. (h) The scalar field $\phi(t)=\phi_0+\int_0^{\frac{1}{2}} \sqrt{\rho+p}~ dt $. Here $n=0.25$}
	
  \label{fig:1}
\end{figure}

If we consider the universe filled of a scalar field $\phi$, then the cosmic density and pressure are related to $\phi$ by
	\begin{equation}
	\rho=K+V ~~~~\text{and} ~~~p=K-V
	\end{equation}
Where $K=\frac{1}{2}\dot{\phi}^2$ is the kinetic term and $V=V(\phi)$ is the potential. This means $K=\frac{1}{2}(\rho+p)$, $V=\frac{1}{2}(\rho-p)$, and $\dot{\phi}^2=(1+\omega) \rho$ with $p=\omega \rho$. The equation of state parameter can be written as 
	\begin{equation} \label{scc}
\omega= \frac{\frac{1}{2}\dot{\phi}^2-V}{\frac{1}{2}\dot{\phi}^2+V} \simeq -1 ~~~ \text{if} ~~~\frac{1}{2}\dot{\phi}^2 \ll V
	\end{equation}
Equation (\ref{scc}) is plotted in Figure \ref{F33}. In other words, when the Kinetic energy of $\phi$ is much smaller than the potential energy, a de Sitter-like period of exponential expansion can be obtained \cite{enqvist}. We can also see that there is an interval where $\omega < -1$. The EoS parameter $\omega$ is directly related to the evolution of $\rho$ and, consequently, to cosmic evolution. If $\omega$ becomes less than $-1$, the energy density $\rho$ will be increasing with cosmic evolution. This can be seen from the conservation equation in FRW cosmology $\dot{\rho}_i=-3H(\rho_i+p_i)$ which implies that $d \ln\rho_i/d \ln a = -3 (1+\omega_i)$. So, $\rho$ would increase as the universe expands if $\omega<-1$. In the literature, considering a scalar field with negative kinetic and gradient energy 
	\begin{equation}
\rho_{\phi}=-\frac{1}{2}\dot{\phi}^2-\frac{1}{2}(\nabla \phi )^2+V(\phi)
	\end{equation}
is the simplest way to get a phantom component where $\omega < -1$ \cite{carolalars}. For the current model, we have for the kinetic term  
\begin{eqnarray}
	K=\frac{\lambda n t_s(2t-t_s)(t-t_s)^{-1}}{4\sqrt{\pi} t\sqrt{\lambda(4\pi \lambda t^4-8\pi \lambda t^3t_s+4\pi \lambda t^2t_s^2-\Lambda t^4+2\Lambda t_s t^3- \Lambda t^2t_s^2+3n^2t_s^2)}}\left\{
        \begin{array}{ll}
            >0 , t \in (0,\frac{1}{2}] \\
            <0 , t \in [\frac{1}{2},1)
        \end{array}
    \right.
	\end{eqnarray}
and for the potential
\begin{eqnarray}  \label{pott}
V &=&  \frac{-\lambda ( t- t_s )^{-1}}{{4 \sqrt \pi t\sqrt {\lambda \left( {4\pi t^4 \lambda  - 8\pi t^3 \lambda t_s  + 4\pi t^2 \lambda t_s^2  - t^4 \Lambda  + 2t^3 \Lambda t_s  - t^2 \Lambda t_s^2  + 3n^3 t_s^2 } \right)}  }} \times \\  \nonumber
&&( 8\pi t^4 \lambda- 16\pi t^3 \lambda t_s  + 8\pi t^2 \lambda t_s^2  - 2t^4 \Lambda  + 4t^3 \Lambda t_s  - 2t^2 \Lambda t_s^2  + 6n^3 t_s^2+ 2nt_s t - nt_s^2 +\\  \nonumber
 &&2t^2 \sqrt \pi  \sqrt {\lambda \left( {4\pi t^4 \lambda  - 8\pi t^3 \lambda t_s  + 4\pi t^2 \lambda t_s^2  - t^4 \Lambda  + 2t^3 \Lambda t_s  - t^2 \Lambda t_s^2  + 3n^3 t_s^2 } \right)}  - 4t\sqrt \pi \\    \nonumber
&&\sqrt {\lambda \left( {4\pi t^4 \lambda  - 8\pi t^3 \lambda t_s  + 4\pi t^2 \lambda t_s^2  - t^4 \Lambda  + 2t^3 \Lambda t_s  - t^2 \Lambda t_s^2  + 3n^3 t_s^2 } \right)} t_s  ) \left\{
        \begin{array}{ll}
            <0 , t \in (0,0.1] \\
            >0 , t \in [0.1,1)
        \end{array}
    \right.
\end{eqnarray}

So, in the current Swiss-cheese Brane-world model, both $V$ and $K$ undergoes a sign change. The potential $V$ changes sign from negative (in the very early epoch) to positive, and the Kinetic term $K$ changes sign from positive to negative. The future big rip singularity occurs with positive $V$ and negative $K$. As we approach the initial singularity ($t \rightarrow 0$), $V$ becomes negative with positive $K$. The expression for the scalar field $\phi(t)$ can be obtained as
\begin{eqnarray} \label{scalarf}
\phi(t)&=& \phi_0+\int_0^1 \sqrt{\rho+p}~ dt \\ \nonumber
&=&\phi_0+\int_0^1 \left(\frac{\lambda n t_s(2t-t_s)(t-t_s)^{-1}}{2 t\sqrt{\lambda \pi (4\pi \lambda t^4-8\pi \lambda t^3t_s+4\pi \lambda t^2t_s^2-\Lambda t^4+2\Lambda t_s t^3- \Lambda t^2t_s^2+3n^2t_s^2)}}\right)^{\frac{1}{2}} ~dt
\end{eqnarray}
Where $\phi_0=\phi(t_0)$. This integration can be numerically evaluated and plotted Taking into account the domain of the function $\sqrt{\rho +p}$ where $\rho+p$ is positive only in the interval $(0,\frac{1}{2}]$ (equation (\ref{null}) for the NEC). We can use $\phi_0=0$ without loss of generality. So, the upper boundary of the integral must be $\frac{1}{2}$ and we obtain an increasing function $\phi(t)$ starting from the initial value $\phi_0$ up to $\frac{1}{2}$. However, obtaining an expression for $t(\phi)$ to substitute in (\ref{pott}) and then plot $V(\phi)$ is difficult for this case. 
\section{The cosmographic expansion}
A major advantage of the cosmographic analysis \cite{cosmography1,cosmography2} is that we can express cosmological parameters only in terms of kinematics. Consequently, the analysis is independent of the particular model considered and it doesn't require an equation of state to probe cosmic dynamics \cite{cosmography3}. The expansion of the scale factor $a(t)$ around the present time $t_0$ is given by
\begin{equation} \label{taylor}
a(t)=a_0 \left[ 1+ \sum_{n=1}^{\infty} \frac{1}{n!} \frac{d^na}{dt^n} (t-t_0)^n \right]
\end{equation}
The following expansion's coefficients are commonly indicated respectively as the Hubble $H$, deceleration $q$, jerk $j$, snap $s$, lerk $l$ and max-out $m$ parameters 
\begin{eqnarray} 
H=\frac{1}{a}\frac{da}{dt}~~,~~q=-\frac{1}{aH^2}\frac{d^2a}{dt^2}~~,~~j=\frac{1}{aH^3}\frac{d^3a}{dt^3}\\   \nonumber
s=\frac{1}{aH^4}\frac{d^4a}{dt^4}~~,~~l=\frac{1}{aH^5}\frac{d^5a}{dt^5}~~,~~m=\frac{1}{aH^6}\frac{d^6a}{dt^6}.
\end{eqnarray} 
It is also possible to relate the derivative of $H$ to the other cosmosmographic parameters as follows \cite{cosmo11}
\begin{eqnarray*} 
\dot{H}&=&-H^2 (1+q)~~,~~\ddot{H}=H^3 (j+3q+2)~~,~~\frac{d^3H}{dt^3}=H^4[s-4j-3q(q+4)-6],\\
\frac{d^4H}{dt^4}&=&H^5[l-5s+10(q+2)(j+3q)+24].
\end{eqnarray*} 
A high-redshift analysis up to the fifth order has been performed in \cite{cosmo11} to constrain the cosmographic expansion. The analysis confirmed the current accelerating cosmic expansion while the estimation of the jerk parameter $j$ shows a possible deviation from the standard $\Lambda CDM$ model. The expressions for $H$ and $q$ for the current BR model have been given in (\ref{dp}), the rest of the parameters are given as
\begin{eqnarray} 
j&=& \frac{1}{n^2t_s^2} \left[ 6t^2+6tt_s(n-1)+t_s^2(n^2-3n+2) \right],\\
s&=& \frac{1}{n^3t_s^3} \left[ 24t^3+36t_st^2(n-1)+12t(n^2t_s^2-3nt_s^2+2t_s^2)+t_s^3(n^3-6(n^2-1)+11n) \right],\\
l&=& \frac{1}{n^4t_s^4} \left[ 120t^4+240t_st^3(n-1)+120t_s^2t^2(n^2-3n+2)+20t_s^3t(n^3-6(n^2-1) \right. \\   
 &+& \left. 11n)+t_s^4(n^4-10n^3+35n^2-50n +24 )\right] \\
m&=& \frac{1}{n^5t_s^5} \left[ 720t^5+1800t_st^4(n-1)+1200t_s^2t^3(n^2-3n+2)+300t_s^3t^2(n^3-6n^2+11n  \right. \\   
 &-&\left. 6) +30t_s^4t(n^4-10n^3+35n^2-50n+24)+t_s^5(n^5-15n^4+85n^3+274n-120)\right]. \nonumber
\end{eqnarray}
The jerk parameter $j$ gives a convenient way to describe models close to $\Lambda CDM$. The parameter $s$ is needed for probing dark energy evolution. The sign change of $q$ also indicates if the expansion is accelerating or decelerating. In spite of the advantages, a detailed discussion of the cosmographic approach's disadvantages and limitations has been presented in \cite{cosmography1}. The evolution of these parameters with cosmic time has been illustrated in figure 2.
\begin{figure}[H] \label{tap10}
  \centering             
  \subfigure[$H$]{\label{F101}\includegraphics[width=0.29\textwidth]{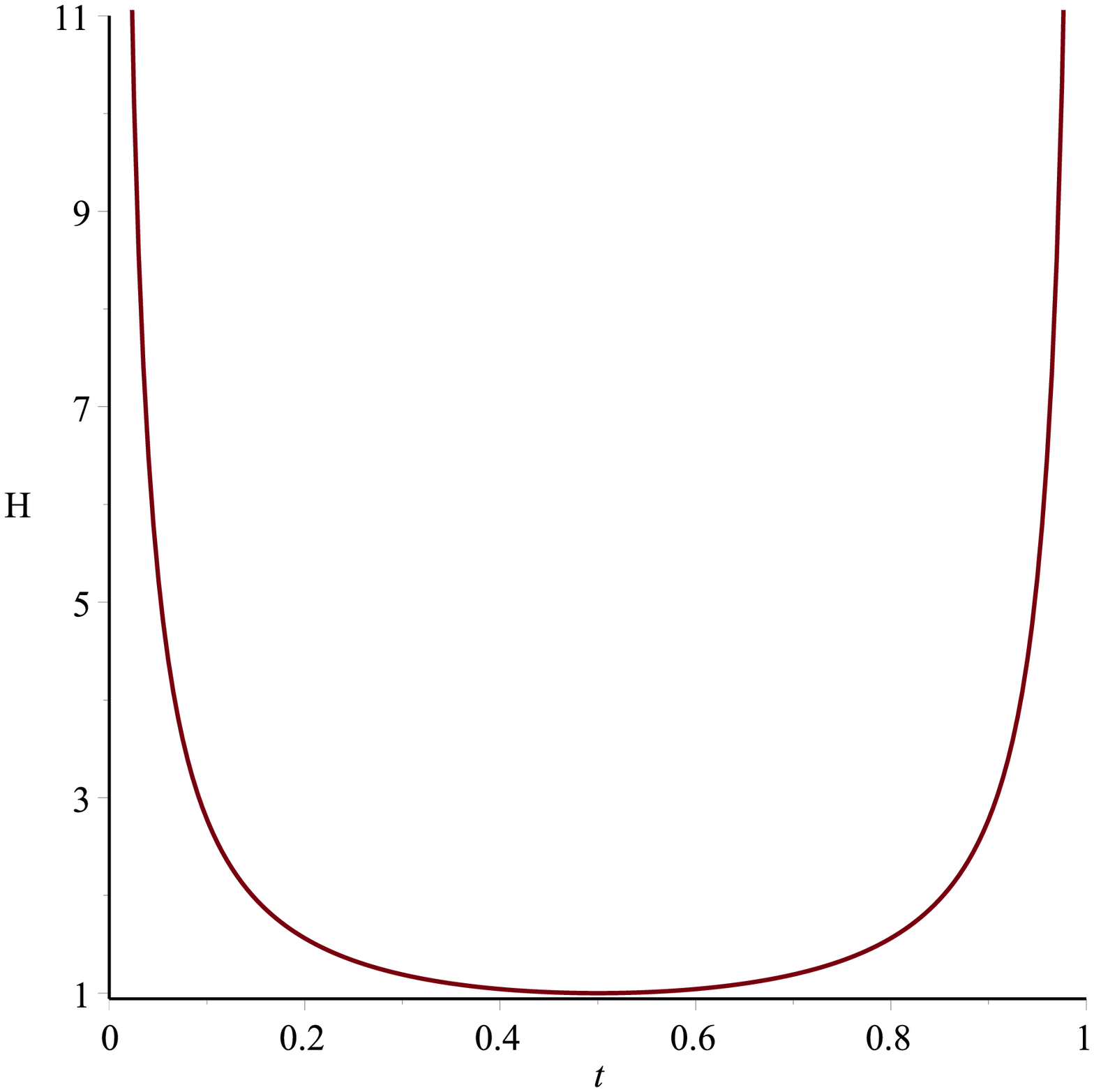}} 
	\subfigure[$q(t)$]{\label{F35}\includegraphics[width=0.29\textwidth]{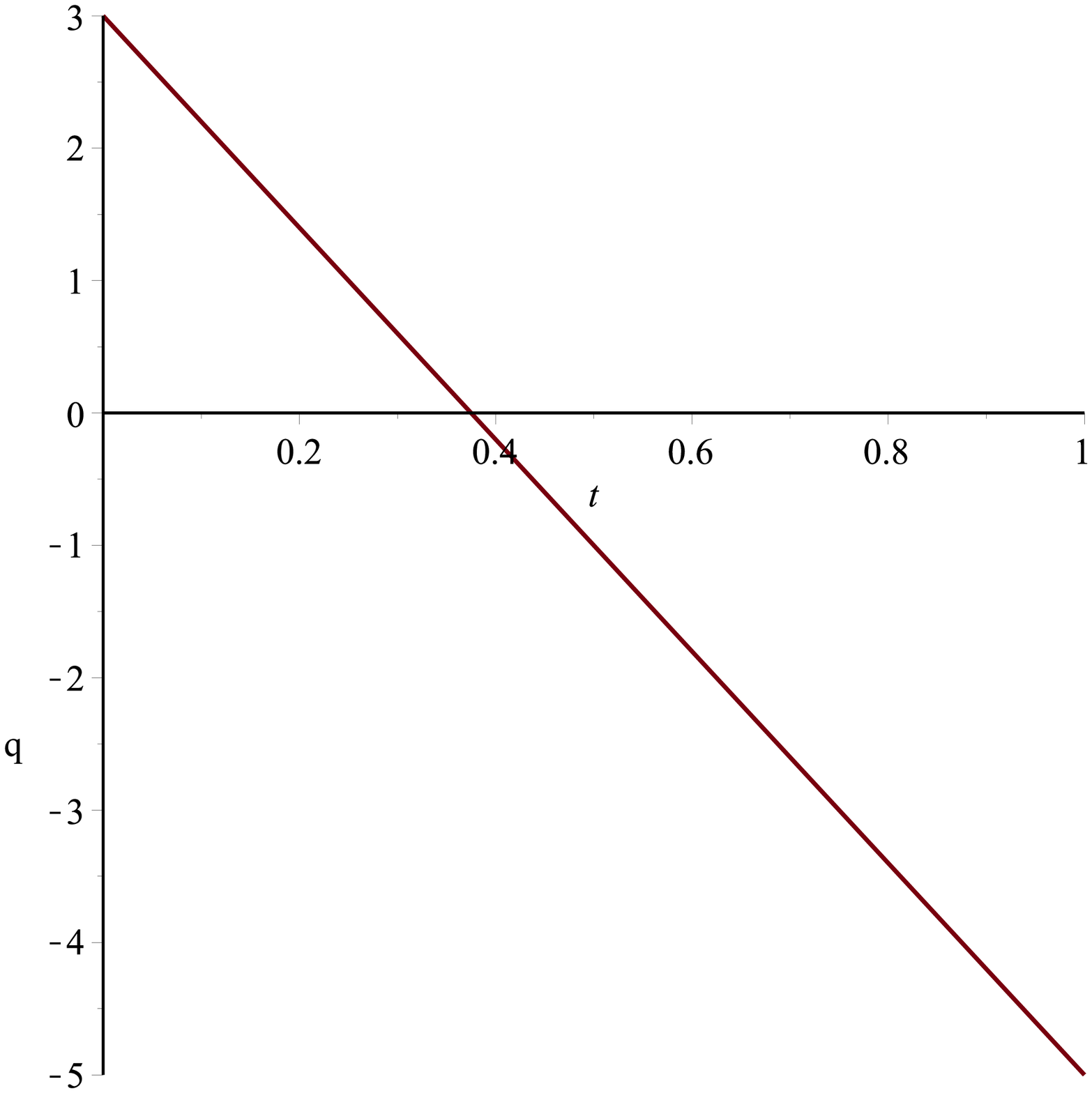}} 
	\subfigure[$j$]{\label{F12}\includegraphics[width=0.29\textwidth]{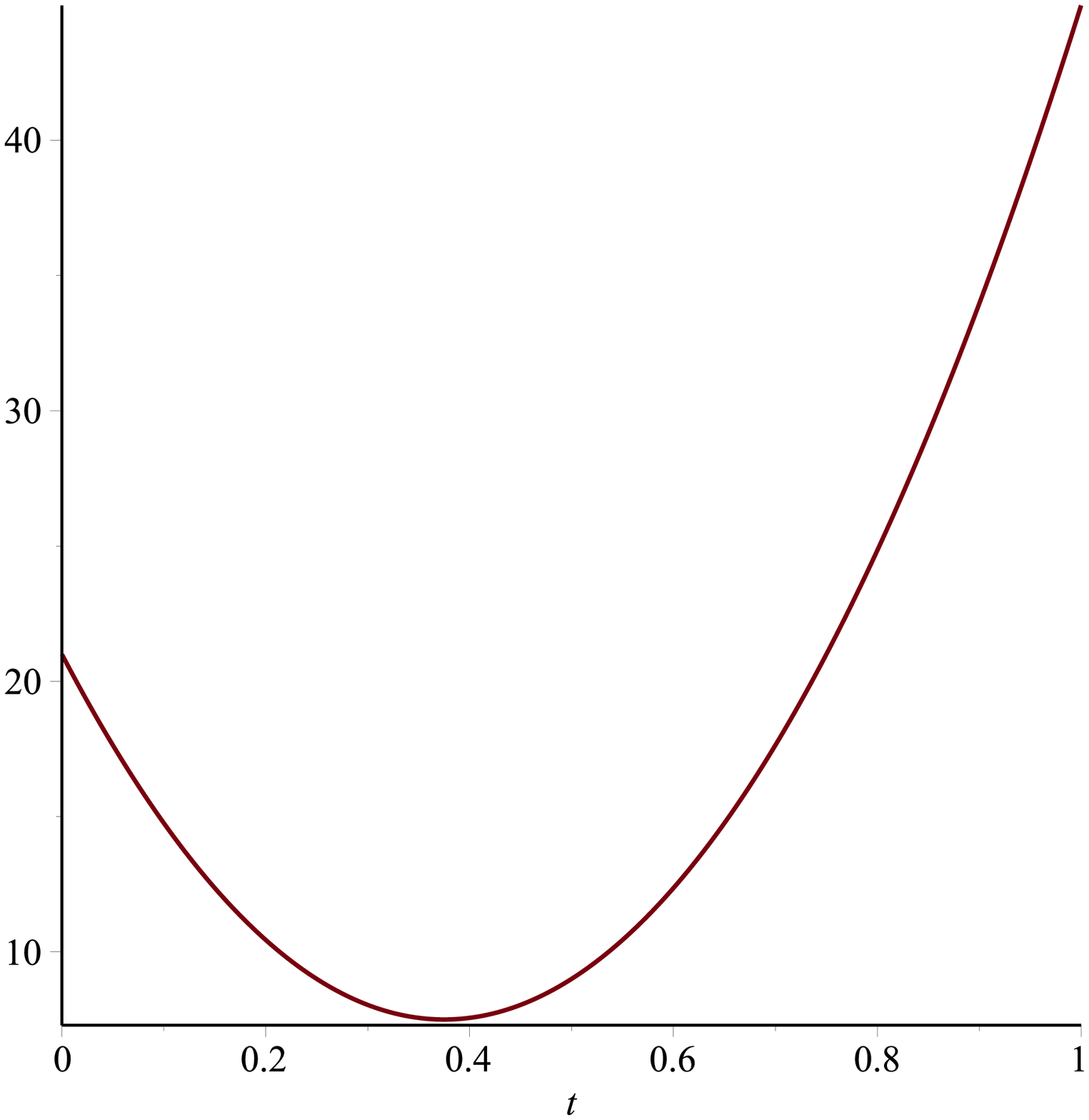}} \\
	\subfigure[$s$]{\label{F13}\includegraphics[width=0.29\textwidth]{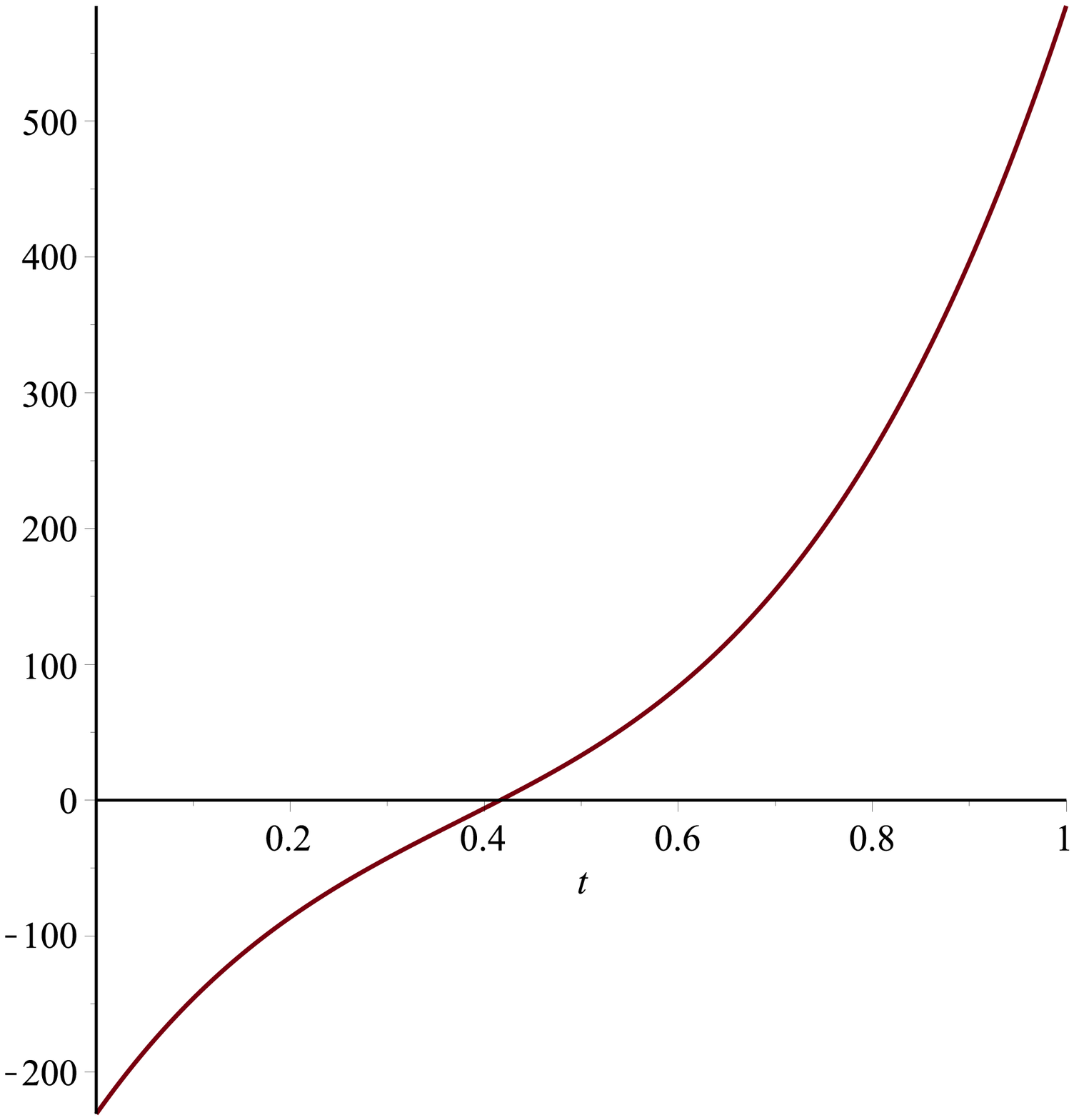}} 
	\subfigure[$l$]{\label{F14}\includegraphics[width=0.29\textwidth]{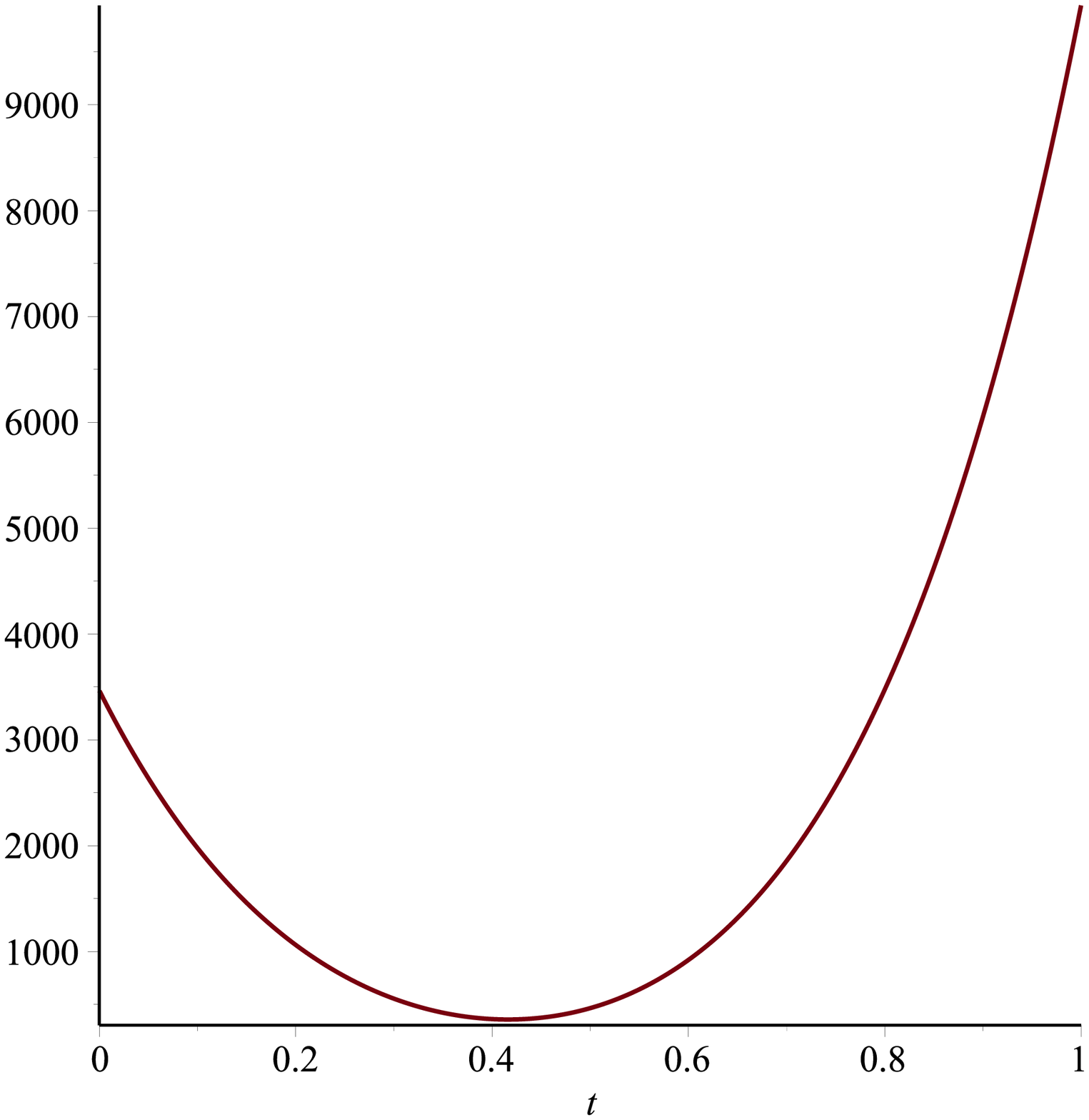}} 
	\subfigure[$m$]{\label{F15}\includegraphics[width=0.29\textwidth]{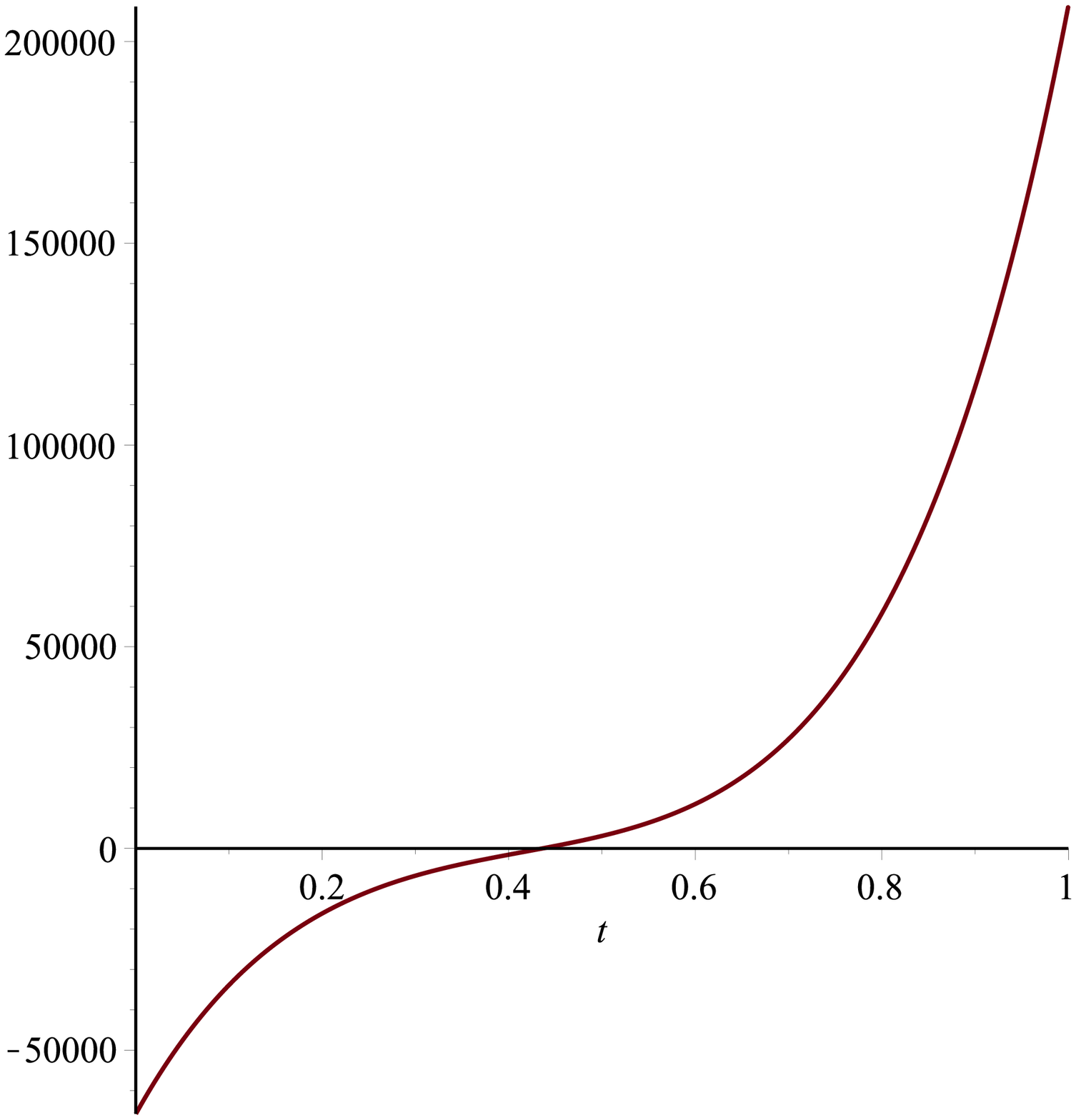}} 
  \caption{Evolution of cosmographic parameters against cosmic time for n=0.25}
  \label{fig:2}
\end{figure}
\section{Conclusion}
We have investigated the big rip future singularity in Swiss-cheese Brane-worlds, the main points of the current work can be summarized as follows:
\begin{itemize}
\item The model is $\Lambda$-independent where the evolution of all parameters has been found to be independent of the value of the cosmological constant $\Lambda$ whether it is $>0$, $<0$ or $=0$. This $\Lambda$-independence of the Swiss-cheese Brane-worlds has been noted before in \cite{mypaper}. All other parameters have been fine-tuned so that $\omega\rightarrow -1$ when $z\simeq 0$ for the current epoch.

\item The expression for the energy density shows no possibility for negative tension branes to exist. It has been found in \cite{negativebr} that a negative tension brane is an unstable object. Also, $\rho$ increases with cosmic expansion for $\omega<-1$.

\item Unlike the original Swiss-cheese brane-world model where the pressure is always positive with no cosmic transit, the pressure in the current model changes sign along with the sign flipping of the deceleration parameter both from $+$ve to $-$ve. This scenario is in agreement with the DE assumption as a component of negative pressure which has a repulsive gravity effect.

\item The evolution of the EoS parameter shows the presence of three phases: the matter dominant decelerating era above $\omega=-1/3$, the accelerated-Quitessence phase between the line $\omega=-1/3$ and the phantom divide line $\omega=-1$, and the phantom phase below the line $\omega=-1$.

\item The violation of the classical linear energy conditions has been illustrated. Due to the existence of the quadratic density term in the cosmological equations, we have also tested the non-linear energy conditions which have been found to be violated too.

\item  For a scalar field $\phi$, the kinetic term $K$ and the potential $V$ both changes sign. The model starts with negative potential (and positive $K$), and ends with positive potential (and negative $K$).

\end{itemize} 
\section*{Acknowledgment}
We are so grateful to the reviewer for his many valuable suggestions and comments that significantly improved the paper. This paper is based upon work supported by Science, Technology \& Innovation Funding Authority (STDF) Egypt, under grant number $37122$.

\section*{Competing interests}
The authors declare there are no competing interests.
\section*{Data Availability}
This manuscript does not report data.

\end{document}